\documentstyle[twocolumn,floats,aps,epsf]{revtex}

\begin{document}

\twocolumn[\hsize\textwidth\columnwidth\hsize\csname@twocolumnfalse%
\endcsname]

{\bf \noindent Comment on ``Bethe Ansatz Results for the 4$f$-Electron
Spectra of a Degenerate Anderson Model ''}

\thispagestyle{empty}

\narrowtext

\vspace{0.5cm}

In a recent letter \cite{Z}, the author calculates  the density 
of states for $4f$ electrons coupled to a conduction band
in the framework of the Bethe ansatz (BA) solution for the degenerate
Anderson model. It is claimed that the results {\it qualitatively 
disagree} with the results obtained
for the same model but using a variational approach \cite {GS}. 
Even the {\it high energy} feature in the $f$-spectral
function near the $4f$-level energy $\epsilon_f$, i.e. the 
``normal'' ionization peak (NIP), is argued to be {\it qualitatively} 
different in the two approaches.
In the following we point out that this is {\it not} the case.

We concentrate on the $U \to \infty$, $N$-fold degenerate Anderson
model. As the Bethe ansatz for this model can yield exact results
only in the large bandwidth limit $B \gg |\epsilon_f| $, we confine 
our discussion to this limit. Zvyagin first presents BA 
results which show that the NIP is shifted towards the chemical
potential from $\epsilon_f$ in the Kondo limit $n_f \approx 1$,
where $n_f$ is the mean $f$-occupancy. He then states that the peak
shifts in the opposite direction in our variational calculation
\cite{GS}. That this is {\it not} the case is obvious from our 
results in the Kondo limit presented in Appendix C of reference \cite {GS}.
This is also shown  very clearly in Fig. $5$ of our
handbook article \cite {HB}. The NIP position is called 
$\tilde \epsilon_f$ in Appendix C of reference \cite {GS}. 
For large $N$ and the Kondo limit this {\it real} quantity
is determined by the equation
\begin{equation}
\tilde \epsilon_f =\epsilon_f +Re \tilde \Gamma ( \tilde \epsilon_f
),
\end{equation}
 where
\begin{equation}
\tilde \Gamma (z)=N\int^0_{-B}d\epsilon |V(\epsilon)|^2/(z-\epsilon).
\end{equation} 
$Re \tilde \Gamma (\epsilon)  $ is {\it positive } for negative
$\epsilon$ with $|\epsilon|\ll B$ (for a constant $V(\epsilon)$ 
it is given by $N\Delta \ln (B/|\epsilon|)$). 
It therefore follows without actually
solving the equation that in our variational calculation the
NIP shifts {\it towards} the chemical potential as in the BA solution.
Zvyagin claims that the solution to this equation was found by us 
``in the {\it complex} form'' \cite{Z}. Apparently he was confused by an
unfortunate misprint in the second equality of our equation which
determines $\tilde \epsilon_f$.  It should read
 $|\tilde \epsilon_f| $ instead
of $\tilde \epsilon_f$ in the argument of the logarithm.
 From the first part of the equation presented
above it should be obvious that this is in fact a misprint.

We should mention that the results for the $f$-spectral function 
using our variational method \cite {GS} essentially agree with the 
low temperature solution of the NCA equations and numerical 
renormalization group results \cite {Hewson}. As noted by Zvyagin\cite {Z}
other low energy characteristics obtained within
our scheme qualitatively coincide with the BA ones.

In conclusion we have pointed out that in contrast to the claims in
reference \cite{Z}  our variational results for the high energy 
feature in the $f$-spectral function for the degenerate Anderson
model, which has been used successfully in the description of
photoemission spectra of Cerium compounds, 
qualitatively {\it agree} with the BA results.

\noindent O.\ Gunnarsson$^1$ and K.\ Sch\"onhammer$^2$\\
$^1$Max-Planck-Institut f\"ur Festk\"orperforschung\\
Heisenbergstrasse1\\
D-70569 Stuttgart\\
Germany\\      
$^2$Institut f\"ur Theoretische Physik\\
Universit\"at G\"ottingen \\
Bunsenstr.\ 9 \\
D-37073 G\"ottingen \\
Germany

\end{document}